# Two-Dimensional Arsenene Oxide: A Realistic Large-gap Quantum Spin Hall Insulator

*Ya-ping Wang,[1] Wei-xiao Ji,[1] Chang-wen Zhang*,[1] Ping Li,[1] Shu-feng Zhang,[1] Pei-ji Wang,[1] Sheng-shi Li,[2] and Shi-shen Yan[2]*

[1] School of Physics and Technology, University of Jinan, Jinan, Shandong, 250022, People's Republic of China

[2] School of Physics, State Key laboratory of Crystal Materials, Shandong University, Jinan, Shandong 250100, People's Republic of China

Searching for two-dimensional (2D) realistic materials able to realize room-temperature quantum spin Hall (QSH) effects is currently a growing field. Here, we through *ab initio* calculations to identify arsenene oxide, AsO, as an excellent candidate, which demonstrates high stability, flexibility, and tunable spin-orbit coupling (SOC) gaps. In contrast to known pristine or functionalized arsenene, the maximum nontrivial band gap of AsO reaches 89 meV, and can be further enhanced to 130 meV under biaxial strain. By sandwiching 2D AsO between BN sheets, we propose a quantum well in which the band topology of AsO is preserved with a sizeable band gap. Considering that AsO having fully oxidized surfaces are naturally stable against surface oxidization and degradation, this functionality provides a viable strategy for designing topological quantum devices operating at room temperature.

One of the grand challenges in condensed matter physics and materials science is to develop dissipationless electron conduction operating at room temperature. The appearance of band topology opens the door to fields of topological quantum states (TQSs), e.g., topological Dirac semimetal[1,2], Weyl semimetal[3,4], and node-line semimetal [5,6], as well as topological insulators (TIs)[7] and so on. Remarkably, two-dimensional (2D) TIs hosts quantum spin Hall (QSH) effect with one-dimensional helical edge states, which can serve as a "two-lane highway" protected by time-reversal symmetry (TRS), making it more suited for coherent spin transport than 3D TIs.[8,9] However, the experimental observations of quantized Hall conductance through QSH effect are only reported in HgTe/CdTe[10,11] and InAs/GaSb[12,13] quantum-wells with a ultralow temperature (<10 K). Hence considerable effort has been devoted to design new materials with QSH effect, and a variety of large-gap 2D TIs have been proposed, including the honeycomb lattices of silicene,[14,15] germanene,[16,17] stanene,[18-21] plumbene,[22] transition-metal halide[23], ZrTe$_5$/HfTe$_5$[24], III-V bilayers[25], BiF[26], and Bi/Sb[27], but none of them has been directly confirmed in experiments. Searching for realistic 2D TIs with large gap that can support high-temperature applications is vitally essential.

* Corresponding author: C. W. Zhang: ss_zhangchw@ujn.edu.cn

Remarkably, the TQSs in oxidized compounds have sparked extensive research interests because of their novel physicochemical properties. The examples available in the literature include a 2D TI in iridate $Na_2IrO_3$[28], topological Mott insulators[29], Weyl semimetals in pyrochlore $Y_2Ir_2O_7$[30], and axion insulators in spinel Osmate[31]. Additionally, the family of oxidized Mxene carbides $M'_2M''C_2$ and $M'_2M_2''C_3$, where $M'$ and $M''$ stands for transition metals has been experimentally synthesized and some of them are identified as 2D TIs.[32,33] The noticeable advantages of oxygen compounds, i.e., naturally antioxidant and stable upon exposure to air, have stimulated continuous efforts in searching for new oxide materials. Recently, the oxidized group-V sheets have been reported as a 2D materials for stretchable display devices, broadband photonic tuning and aberration-free optical imaging.[34] Motivated by the great interest in search of nontrivial TIs experimentally, questions naturally arise: whether the 2D asrenene oxide, AsO, can host QSH effect, and if yes, it can be used at high temperature without dissipation which makes it potentially suitable to macroelectronics?

In this work, based on density-functional theory (DFT) calculations, we propose a new intrinsic QSH insulator in 2D AsO, where oxygen is the surface functional group. Due to the orbital filtering effect (OFE) of oxidation, the SOC on $p_{x,y}$ orbitals of As atom is enhanced significantly with its band gap reaching 89 meV, and can be further enhanced to 130 meV under biaxial strain. By sandwiching AsO between BN sheets, the BN/AsO/BN quantum well remains topologically nontrivial with a sizeable band gap. Considering that 2D AsO having fully oxidized surfaces is naturally stable against surface oxidization and degradation, our findings present an ideal platform for realizing QSH effect in spintronics devices.

First-principles calculations are performed by using DFT methods as implemented in the Vienna *ab initio* simulation package (VASP).[35] The projector-augmented -wave (PAW) potential[36,37], Perdew-Burke-Ernzerhof (PBE) exchange-correlation functional[38] and the plane-wave basis with a kinetic energy cutoff of 500 eV are employed. The Brillouin zone is sampled by using an 11×11×1 Gamma-centered Monkhorst–Pack grid. The vacuum space is set to 30 Å to minimize artificial interactions between neighboring slabs. During structural optimization, all atomic positions and lattice parameters are fully relaxed, and the maximum force allowed on each atom is less than 0.02 eV/ Å. SOC is included by a second variational procedure on a fully self-consistent basis. Furthermore, the screened exchange hybrid density functional by Heyd-Scuseria-Ernzerhof (HSE06)[39] is adopted to check the electronic structures. The phonon calculations are carried out by using DFT perturbation theory as implemented in the PHONOPY code[40] combined with VASP.

We consider two different films of arsenene oxide: (i) Dandruff comb (a double-edged fine-toothed comb) hereinafter referred to as D-AsO, characterized by oxygen atoms alternating on both sides of the sheet, as shown in Fig. 1(a); and (ii) Comb shape is referred to as C-AsO, where the oxygen atoms are at the same side of the sheet in Fig. 1(b). The calculated equilibrium lattice constants are 4.53 and 4.94 Å for D-AsO and C-AsO, respectively, while the corresponding buckling parameters are 0.95 and 0.00Å. The As-As-As bond angles are 108.5 °and 120.1 °, corresponding to O-As-As bond angles of 110.4 ° and 90.1 ° for D-AsO and C-AsO. These angles changes indicates, upon oxidization, the formation of a buckled honeycomb lattice for D-AsO is similar with the case of transition from silicene to silicane due to the $sp^3$ hybridization.[41,42] Figure 1(c) presents the total energy per unit cell as a function of lattice constant *a*. Interestingly, it displays two local minima in energy of D-AsO, which is defined as low-buckled and high-buckled structures. The most stable phase is low-buckled phase while the high-buckled does not meet the scope of realistic research.

However, the C-AsO structure is different from D-AsO, and it has only one stable phase.

The energetically stability of both these configurations can be evaluated by calculating the formation energy expressed as $\Delta E_f = E(\text{AsO}) - E(\text{arsenene}) - \mu(\text{O}_2)$, where $E(\text{AsO})$ and $E(\text{arsenene})$ are the total energies of AsO and pristine arsenene, respectively, while $\mu(\text{O}_2)$ is the chemical potential of oxygen gas. The results for two isomeric sheets indicates that both D-AsO (-7.90 eV) and C-AsO (-7.11 eV) are energetically more stable than the hydrogenated or halogenated ones.[43,44] To further address their dynamic stability, we perform the phonon spectrum dispersion calculations and the results are presented in Figs. 1(d) and (e). It is clearly seen that D-AsO is rather stable because all the vibrational modes are real in the whole Brillouin zone, whereas C-AsO are dynamically unstable under ambient condition due to the appearance of negative vibrations modes at the K point.

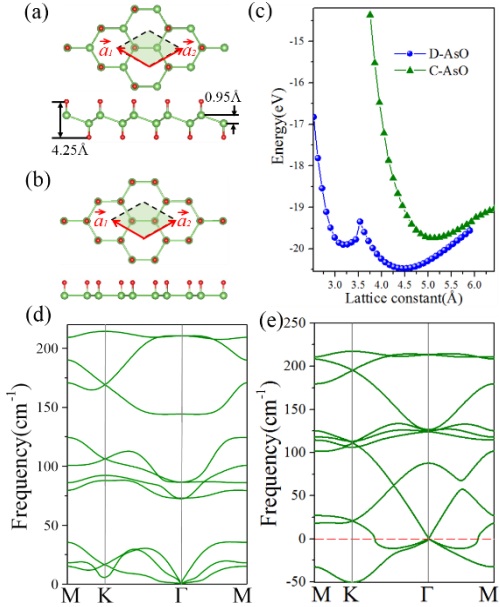

**Fig. 1** Structural representation (top view and side view) of (a) D-AsO and (b) C-AsO. (c) Total energy with respect to lattice constant of D-AsO and C-AsO (d) and (e) correspond to phonon spectra along the high-symmetric points in the BZ.

As is known, functionalization will dramatically affect the electronic and magnetic properties of 2D materials. We thus calculate the band structure of D-AsO and C-AsO isomers, as show in Fig. 2. In the absence of SOC, The conduction band minimum (VBM) and valence band maximum (VBM) touch each other at the Γ point, indicating both systems are essentially gapless. From the 3D plot of band dispersions around the Fermi level in the insert of Figs. 2(a) and (b), one can see that that the valence and conduction bands at the Γ point are both parabolic. However, when SOC is taken into account in the calculations, it leads to a large band gap (89 meV for D-AsO and 86 meV for C-AsO). Such a SOC-induced metal-insulator transition strongly suggests that they are 2D TIs. Since the PBE method usually underestimates the band gap, we examine the band gaps using HSE06,[39] and find that the similar band topologies remains with their band gaps being enhanced to 232 meV, comparable with the ones of functionalized buckled Bi (111) (0.2 eV),[27] $ZrTe_5$ (0.1 eV)[24], as well as stanene (0.1 eV)[18-21]. These sizeable bulk gaps can stabilize the edge states against the interference of the thermally activated carriers, which is beneficial for observing room-temperature QSH effect.

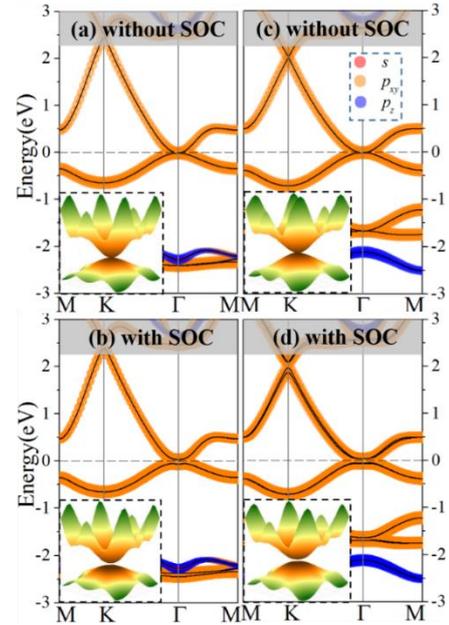

**Fig. 2** The band structures of D-AsO (a and b) as well as C-AsO (c and d) without (a and c) and with (b and d) SOC, weighted with $s$, $p_{x,y}$ and $p_z$ character, respectively.

The insert is the corresponding 3D band structures for both structures.

A hallmark of 2D TI is the presence of an odd number of topologically helical edge states that cross the Fermi level, protected by TRS. To see these features explicitly, we calculate the edge Green's functions of a semi-infinite D-AsO using *ab initio* tight-binding (TB) model, with the recursive method based on maximally localized Wannier functions (MLWFs).[45] Using D-AsO as a representative, in Figures 3(c) and (d) we displays the fitted DFT and TB model bands, as well as local density of states (LDOS) on the edge states. One can explicitly see that each edge has a single pair of helical edge states in the bulk band gap and cross linearly at $\Gamma$ point. Besides, by identifying the spin-up (↑) and spin-down (↓) contributions in edge spectral function, the counter-propagating edge states exhibit opposite spin-polarization, making them compatible with the spin-momentum locking of 1D helical electrons. These nontrivial topology can further be confirmed by the calculated $Z_2$ invariant, obtained to trace the Wannier Center of Charges (WCCs) using non-Abelian Berry connection.[46] By counting the number of crossings of any arbitrary horizontal reference line, the WCCs evolution curves cross any arbitrary reference lines odd times, yielding $Z_2 = 1$, illustrated in Fig. 3(e).

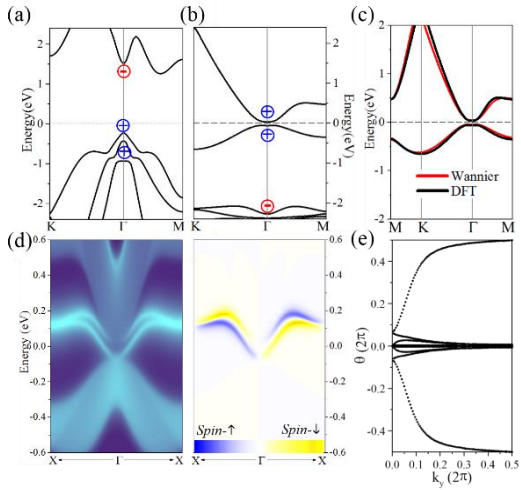

**Fig. 3** Band structure for (a) arsenene, (b) arsenene oxide. (c) DFT and MLWFs fitted band structures for D-AsO. (d) The calculated helical edge states of D-AsO, where the left subpanel shows the total DOS while the right subpanel shows the corresponding spin polarization in both spin up and spin down channels. (e) Evolutions of Wannier centers along $k_y$, yielding $Z_2 = 1$.

To better understand why AsO has a large band gap, we next do an orbital analysis around the Fermi level in Fig. 2, in which the red, orange and blue colors stand for the projected s, $p_{x,y}$ and $p_z$ bands with different orbital symmetries, respectively. One can see that the decorated oxygen atoms hybridizes strongly with the dangling bonds of As-$p_z$ orbital, which effectively removes the $p_z$ bands away from the Fermi level, leaving only $p_{x,y}$ states at the Fermi level. This effect can be attributed to the known orbital filtering effect (OFE), proposed in the epitaxial growth of Bi on halogenated Si(111) substrate. Hence the orbital evolution of D-AsO at the $\Gamma$ point is shown in Fig. 4(b) as a representative. Obviously, the chemical bonding between As-As atoms make $p$ states split into the bonding and antibonding states, $|p_{xy}^+\rangle$ and $|p_{xy}^-\rangle$, where the superscripts + and − represent the parities of corresponding states, respectively. In the absence of SOC, the Fermi level cross crosses $|p_{xy}^+\rangle$ states, forming a semi-metallic character, as illustrated in Fig. 2(a). Switching on SOC would make the $|p_{xy}^+\rangle$ orbitals split into $|p_{x+iy,\uparrow}^+\rangle$ & $|p_{x-iy,\downarrow}^+\rangle$ and $|p_{x-iy,\uparrow}^+\rangle$ & $|p_{x+iy,\downarrow}^+\rangle$, lowering $|p_{x-iy,\uparrow}^+\rangle$ & $|p_{x+iy,\downarrow}^+\rangle$ and raising $|p_{x+iy,\uparrow}^+\rangle$ & $|p_{x-iy,\downarrow}^+\rangle$, thus opens a band gap of 89 meV near the $\Gamma$ point. Noticeably, the SOC is essential to open a finite band gap but does not induce the band inversion, which is clearly different from the conventional 2D TIs induced by band inversion mechanism.[18-21] In fact, the similar band characteristic has also observed in SnF monolayer, where the nontrivial band mechanism can be attributed to reversal of parity.[18] As a consequence, it is the SOC from $p_{x,y}$ orbital that the nontrivial band gap are enhanced significantly to a sizable band gap , in comparison to the pristine arsenene[47].

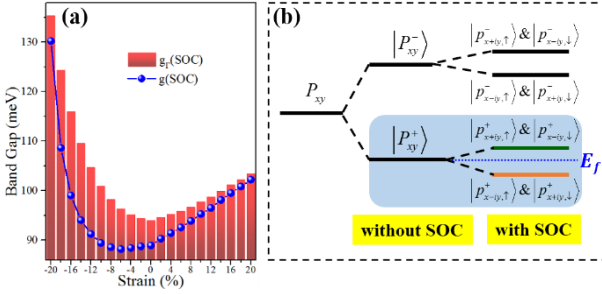

**Fig. 4** (a) The variations of band gap with respect to external strain. (b) The schematic of band evolution at the Γ point, the Fermi levels are at energy zero.

It is known that materials composed by the light elements, such as carbon and silicon, have very small SOC strength, rendering them unsuitable to support the QSH effect at room temperature. Hence, to design a large gap QSH insulator, 2D honeycomb lattices from heavy elements such as Hg, Bi, and Pb are expected to ensure strong SOC and achieve large-gap QSH devices.[22,26,27] However, the present works demonstrate that it is plausible to use light elements to form a large-gap QSH insulator *via* OFE. Enlightened by this new insight, the honeycomb lattice made of heavy elements is not necessary. This fact will bring out new strategies for designing 2D TQS materials.

After confirming the 2D TI phases in AsO, we further check the robustness of their nontrivial topologies against external strain. Figure 4(a) shows the evolution of direct gap ($E_Γ$) at Γ point and whole indirect band-gap ($E_g$) as a function of strain $ε$, which is defined as $(a-a_0)/a_0$, where $a$ ($a_0$) is the strained (equilibrium) lattice constants for D-AsO configuration. One can see that the topological invariant $Z_2$ is always within the strain range of ±20 %. This suggests that 2D AsO maintains a topologically nontrivial state, which is stable against strain. Interestingly, a larger whole gap of 130 meV is obtained for compressive strain 120 % in PBE method, and importantly, this large band gap can reach 272 meV in HSE06 functional here. It is noticeable that, though the bulk gap is enhanced for HSE06 method, the $p_{x,y}$ band character is not altered at the Fermi level, indicating that the nontrivial topology is robust to calculated method.

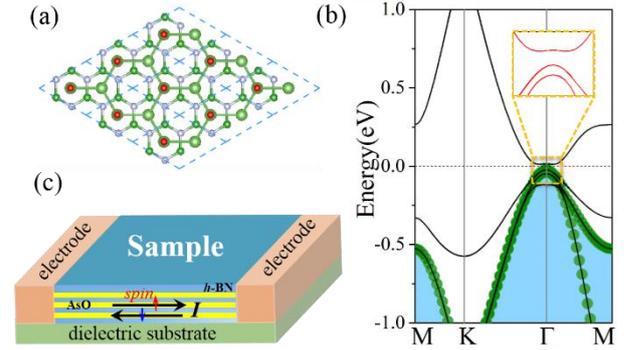

**Fig. 5** Crystal structures of QW consisting of AsO monolayer sandwiched *h*-BN sheet on top view for (a), as well as the corresponding band structure with SOC under 6% tensile strain for (b). The green circles and blue areas represent the size and extent of the substrate effect (c) Schematic model for proposed BN/AsO/BN heterostructure for quantum state measurement. Vertical arrows show the spin orientation of electrons in the edge states and horizontal arrows show their transport directions.

From the material standpoint, it is highly feasible to experimentally realize the epitaxial growth of AsO sheet thanks to the existing related experiments. Fig. 5(a) illustrates the proposed geometrical structures of AsO on $\sqrt{3} \times \sqrt{3}$ BN substrate, where the lattice mismatch is only about 4.96%. After full relaxation with van der Waals (vdW) forces,[48] the D-AsO sheet retains the original structure with a distance of 3.45 Å for BN sheet. The calculated binding energy of this heterostructure is -9.18 meV/atom, showing that it is typical vdW structure.

Figure 5(b) shows the band structure of heterostructure under 6 % tensile strain. As expected, in these weakly coupled systems, there is a SOC-induced band gap opening around the Fermi level, dominantly contributed by the AsO sheet. Considering the wide-gap BN sheet electrically insulate adjacent QSH layers of 2D AsO, protecting parallel helical edge channels from being gapped by interlayer hybridization, the predicted BN/AsO/BN heterostructure can parametrically increase the number of edge transport channels to support the dissipationless charge/spin transport in the topological states. These results demonstrate the possibility of designing quantum transport devices with this heterostructure, as illustrated in Fig. 5(c).

In summary, based on first-principles calculations, we predict an intrinsic large-gap QSH insulator in the realistic AsO sheet. By oxidation to As atoms, the out-of-plane $p_z$ is filtered from $p$ orbitals, thus the strength of SOC on As-$p_{x,y}$ orbitals is enhanced significantly. A single pair of topologically protected helical edges is established, and its QSH phase is confirmed with $Z_2 = 1$. Furthermore, by sandwiching AsO between BN sheets, the BN/AsO/BN quantum well remains topologically nontrivial with a sizeable band gap, suggesting the robustness of its band topology against the effect of the substrate. These findings demonstrate that 2D AsO may be an innovative platform for QSH device design and fabrication operating at room temperature.

Acknowledgments: This work was supported by the National Natural Science Foundation of China (Grant Nos.11434006 and 11304121).